\documentclass[12pt,aps,prb,preprint]{revtex4}   

\usepackage{amsmath}    
\usepackage{graphicx}   
\usepackage{amsfonts}
\usepackage{amssymb}
\usepackage{bm}
\usepackage{bbm}
\usepackage{fancyhdr}
\usepackage{textcomp}
\usepackage{mathrsfs}
\usepackage{enumerate}
\usepackage{color}

\begin{document}

\title{
Combinatorics in tensor integral reduction
}
\author{June-Haak Ee}
 \affiliation{Department of Physics, Korea University,
 Seoul 02841, Korea}
\author{Dong-Won Jung}
 \affiliation{Department of Physics, Korea University,
 Seoul 02841, Korea}
\author{U-Rae Kim}
 \affiliation{Department of Physics, Korea University,
 Seoul 02841, Korea}
\author{Jungil Lee}
 \email{jungil@korea.ac.kr}   
 \affiliation{Department of Physics, Korea University,
 Seoul 02841, Korea}
\date{\today}

\begin{abstract}
We illustrate a rigorous approach to express the totally symmetric isotropic tensors of arbitrary rank in the
 $n$-dimensional Euclidean 
space as a linear combination of products of Kronecker deltas.
By making full use of the symmetries, one can greatly reduce
the efforts to compute
cumbersome angular integrals into 
straightforward combinatoric counts.
This method is generalized into the cases in which such symmetries are
present in subspaces.
We further demonstrate
the mechanism of the tensor-integral reduction that is
widely used in various physics problems
such as perturbative calculations of the gauge-field theory
in which divergent integrals are regularized in $d=4-2\epsilon$ 
space-time dimensions.
The main derivation is given in the $n$-dimensional Euclidean space.
The generalization of the result
to the Minkowski space is also discussed 
in order to provide graduate students and researchers
with techniques
of tensor-integral reduction for particle physics problems.
\end{abstract}

\maketitle
\section{Introduction}
As geometric objects generalized from scalars and vectors, 
tensors act an important role to represent various physical quantities.
The fundamental classification to distinguish scalars, vectors,
and tensors is based on their transformation properties 
under rotation.\cite{Battaglia-2013}
A scalar is a single invariant quantity under rotation.
In the $n$-dimensional Euclidean space, a vector 
$\bm{V}=(V^1,\cdots,V^n)$
has $n$ Cartesian components and a vector index $i$ is used to 
identify the $i$th component $V^i$.
Under a rotation, 
the corresponding linear transformation matrices of any vectors
are identical to a single rotation matrix $\mathscr{R}^{ij}$ 
that transforms a radial vector $\bm{r}=(x^1,\cdots, x^n)$ into
$\bm{r'}=(x^{\prime 1},\cdots, x^{\prime n})$ like
$x^{\prime i}=\sum_{j=1}^n\mathscr{R}^{ij}x^j$
keeping the magnitude $|\bm{r}|=\sqrt{\sum_{j=1}^n(x^j)^2}$ invariant. 
This relation also holds for any polar vector $\bm{V}$.
Then, the direct product of 
$k$ polar vectors, $\mathcal{V}^{i_1\cdots i_k}
\equiv V^{i_1}\cdots V^{i_k}$ must transform like
$\mathcal{V}^{\,\prime i_1\cdots i_k}=
\sum_{j_1=1}^{n}
\cdots\sum_{j_k=1}^{n}
\mathscr{R}^{i_1 j_1}
\cdots \mathscr{R}^{i_k j_k}\mathcal{V}^{j_1\cdots j_k}$.
An object with $k$ vector indices that transforms like 
$\mathcal{V}^{i_1\cdots i_k}$
is called a rank-$k$ Cartesian tensor 
$\mathcal{T}^{i_1\cdots i_k}_{(k)}$.
If a tensor is invariant under rotation, then we call it
 an \textit{isotropic} tensor.
A tensor is called \textit{symmetric} (\textit{antisymmetric}) if it is invariant
(flips the sign)
under exchange of two given vector indices.
Elementary examples of symmetric rank-2 tensors are
the tensor of polarizability, tensor of inertia,
and tensor of stress.
A cross product of two polar vectors such as the angular momentum 
or torque is an antisymmetric rank-2 tensor.
\cite{Feynman-1} 
If a tensor is invariant (flips the sign) under exchange of
any two vector indices, then we call it \textit{totally symmetric}
(\textit{totally antisymmetric}).

Among various tensors, the totally symmetric isotropic tensors
$\widetilde{\mathcal{I}}^{i_1\cdots i_k}_{(k)}$ are
particularly important because of their invariance properties:
$\widetilde{\mathcal{I}}^{i_1\cdots i_k}_{(k)}$ remains
the same under rotation and/or
exchange of any two vector indices. The most elementary
example is the Kronecker delta $\delta^{ij}$ which is of rank 2.
Because of the rotational symmetry of $\delta^{ij}$,
it is trivial to verify that the scalar product
$\bm{A}\cdot\bm{B}$ of any two vectors $\bm{A}$ and $\bm{B}$
is invariant under rotation. If $\delta^{ij}$ is multiplied to
a rank-2 tensor $\mathcal{T}^{ij}_{(2)}$ 
and the vector indices are contracted, then
one obtains the trace of the tensor which is invariant under
rotation.

A considerable amount of research on isotropic tensors to
study various physics problems has been made
such as in
crystallography or rheology describing material properties, \cite{Smith-1968,Smith-1970,Smith-1971,Kearsley-1975}
fluid dynamics to study turbulence effects,
\cite{Hinze-1975, Robertson-1940, Champagne-1978}
and molecular dynamics involving multi-photon 
emission/absorption processes.\cite{Kielich-1961,%
Healy-1975,
Andrews-1977,Andrews-1980}
All of the approaches listed above 
are restricted to three dimensions.
In fact, dimensional regularization in
the gauge-field theory of particle physics requires
information of
$\widetilde{\mathcal{I}}^{i_1\cdots i_k}_{(k)}$
in $d=4-2\epsilon$ space-time dimensions, where $\epsilon$ is
an infinitesimally small number. In this approach
a divergent loop integral appearing in perturbative expansions
is regularized into powers of $1/\epsilon$.
Eventually, after a proper
renormalization, the resultant 
physical quantities become finite as $\epsilon\to 0$
and the theory retains the predictive power.
\cite{HV-1972,Bollini-1972-NC,Bollini-1972-PLB,H-1973-1,H-1973-2}
In employing dimensional regularization,
physical observables are first calculated in $n=d-1$ spatial
dimensions
assuming that $n$ is a countable number.
Then, any functions of $n$, such as the
gamma function, appearing in
physical variables are analytically continued 
to $n=3-2\epsilon$. 
The loop integrals are integrated 
over a loop momentum $p=(p^0,p^1,\cdots,p^{d-1})$ whose every component 
runs from $-\infty$ to $\infty$. Here, $p$ is a $d$-vector,
which is the $d$-dimensional analogue of the four-vector 
$p=(p^0,p^1,p^2,p^{3})$.
After evaluating the residue by integrating over the energy component
$p^0$, the resultant integrals over $p^i$'s involve
angle averages.
This integral is in general a linear combination of 
tensor integrals that are to be simplified into a linear combination
of constant tensors with the coefficients proportional to 
scalar integrals. This procedure is consistent with 
the standard approach called the Passarino-Veltman 
reduction.\cite{Pa-Ve-1979}

In this paper, we
illustrate a systematic approach
to find the explicit form of the totally symmetric isotropic tensor
$\widetilde{\mathcal{I}}^{i_1\cdots i_k}_{(k)}$ as a linear
combination of Kronecker deltas.
 In principle, the expression can be found by taking the average of 
$\hat{{r}}^{i_1}\cdots\hat{{r}}^{i_k}$ over the angle of a unit
radial vector $\hat{\bm{r}}$. For example, 
$\widetilde{\mathcal{I}}^{ij}_{(2)}=\delta^{ij}/3$ in 3 
dimensions.\cite{Zelevinsky-2011}
While a direct evaluation of the average over
$n-2$ polar angles and an azimuthal angle in $n$ dimensions
requires great efforts to deal with quite a few beta functions,
our rigorous derivation based on only the abstract algebraic structure
makes extensive use of the symmetries leading to a 
great simplification of the evaluation steps into simple
counts of combinatorics. Our methods are generalized into
the cases in which such symmetries are present in subspaces.
As applications, we demonstrate how to make use of 
$\widetilde{\mathcal{I}}^{i_1\cdots i_k}_{(k)}$ in evaluating
angular integrals and in tensor-integral reductions.

This paper is organized as follows.
In Sec.~\ref{sec:def}, we list definitions of fundamental
terminologies of tensor analyses that are frequently used
in the remainder of this paper.
Section~\ref{sec:ans} provides the derivation of
the explicit form of the totally symmetric
isotropic tensor $\widetilde{\mathcal{I}}^{i_1\cdots i_k}_{(k)}$.
Applications of $\widetilde{\mathcal{I}}^{i_1\cdots i_k}_{(k)}$ to 
the computation of angular integrals and the tensor-integral reduction
are given in Sec.~\ref{sec:app} which is followed by a summary in 
Sec.~\ref{sec:sum}.
Appendices provide technical formulas: A standard parametrization
of the spherical polar coordinates in the $n$-dimensional Euclidean space
is given in Appendix~\ref{app:polar-azimuthal}.
Explicit evaluations of angle averages
in $n$ dimensions are listed
in Appendix~\ref{app:explicit-evaluation-adotkton}.
The extension of our results to problems in the
$d$-dimensional Minkowski space is
summarized in Appendix~\ref{app:cov}.
\section{Definitions\label{sec:def}}
In this section, we list definitions of terminologies
involving the vector and tensor
analyses presented in this paper.
We work in the $n$-dimensional Euclidean space
$\mathfrak{V}_{(n)}$.
\subsection{Vector}
A vector $\bm{V}\in\mathfrak{V}_{(n)}$ can be expressed
as a linear combination 
\begin{equation}
\bm{V}=V^i\hat{\bm{e}}_{i},
\end{equation}
where $\hat{\bm{e}}_i$ and $V^i$ are 
the unit basis vector along the $i$th Cartesian axis
and
the $i$th component, respectively.
The $n$-tuple $(V^1,\cdots,V^n)$
is also used to denote $\bm{V}$.
Every Cartesian axis of $\mathfrak{V}_{(n)}$ is homogeneous and isotropic
so that the unit basis vectors satisfy the orthonormal conditions: 
\begin{equation}
\hat{\bm{e}}_i\cdot\hat{\bm{e}}_j
=\delta^{ij},
\end{equation}
where the Kronecker delta $\delta^{ij}$
is the $ij$ element of the $n\times n$ identity 
matrix $\mathbbm{1}$.
Thus the scalar product of two vectors
$\bm{V}$ and $\bm{W}$ can be expressed as
\begin{equation}
\bm{V}\cdot\bm{W}=V^iW^i.
\end{equation}
Here and in the remainder of this paper we use
the Einstein's convention for summation of repeated
vector indices: 
$V^iW^i$ represents $\sum_{i=1}^nV^{i}W^{i}$.
The square and the magnitude of a vector $\bm{V}$ are defined by
$\bm{V}^2\equiv\bm{V}\cdot\bm{V}$
and
$|\bm{V}|\equiv \sqrt{\bm{V}^2}$, respectively, and
$\bm{V}^{2k}\equiv(\bm{V}^2)^k$
for any positive integer $k$.
The unit vector $\hat{\bm{V}}$
along the direction of $\bm{V}$ is defined by 
$\hat{\bm{V}}\equiv\bm{V}/|\bm{V}|$.
The trace of the identity matrix
is the dimension of $\mathfrak{V}_{(n)}$:
\begin{equation}
\delta^{ii}=n.
\end{equation}

Under rotation, a vector $\bm{V}$ transforms into $\bm{V}'$
as
\begin{equation}
\label{eq:vector-rotation}
V^{\prime i}=\mathscr{R}^{ij}V^j,
\end{equation}
where $\mathscr{R}$ is the rotation matrix.
Because $\bm{V}^{\prime2}=\bm{V}^{2}$,
$\mathscr{R}$ is an orthogonal matrix:
\begin{equation}
\label{eq:orthogonal}
\mathscr{R}^T\mathscr{R}=\mathscr{R}\mathscr{R}^T=\mathbbm{1}.
\end{equation}

\subsection{Gram-Schmidt orthogonalization}
A convenient way of constructing orthonormal basis vectors is
Gram-Schmidt orthogonalization.\cite{Gram}
If  $\bm{a}_1$, $\cdots$, $\bm{a}_m$ are linearly independent
vectors in $\mathfrak{V}_{(n)}$ with $m\le n$, then one can
construct unit basis vectors as
\begin{equation}
\label{def:eq-from-p=1tom}
\hat{\bm{e}}_p
=\frac {1}{\sqrt{D_{(p-1)}D_{(p)}}}
\begin{vmatrix}
\bm{a}_1&
\bm{a}_2&
\cdots &
\bm{a}_p
\\
\bm{a}_1\cdot\bm{a}_1&
\bm{a}_2\cdot\bm{a}_1&
\cdots&
\bm{a}_p\cdot\bm{a}_1
\\
\vdots&\vdots&\ddots&\vdots
\\
\bm{a}_1\cdot\bm{a}_{p-1}&
\bm{a}_2\cdot\bm{a}_{p-1}&
\cdots&
\bm{a}_p\cdot\bm{a}_{p-1}
\end{vmatrix},
\end{equation}
where $p=1,$ $\cdots$, $m$,
$D_{(p)}=\textrm{det}[A_{(p)}]$
is called
the Gram
determinant of a $p\times p$ square matrix $A_{(p)}$ with elements
$A_{(p)}^{ij}=\bm{a}_i\cdot\bm{a}_j$,
and $D_{(0)}=1$.
\subsection{Tensor}
The symbol
$\mathcal{T}^{i_1\cdots i_k}_{(k)}$ denotes
the  Cartesian tensor of rank $k$, where $k$ is the number of vector indices.
Under rotation,
$\mathcal{T}^{i_1\cdots i_k}_{(k)}$ transforms like
\begin{equation}
\mathcal{T}^{\prime\,i_1\cdots i_k}_{(k)}=
\mathscr{R}^{\,i_1j_1}
\cdots
\mathscr{R}^{\,i_kj_k}
\mathcal{T}^{j_1\cdots j_k}_{(k)}.
\end{equation}
\subsection{Permutation\label{sec:perm}}
In the remainder of this paper, we use the
 symbol $\sigma$ to represent one of the $k!$ permutations
$[\sigma(i_1),\cdots,\sigma(i_k)]$ of the
ordered list of vector indices $(i_1,\cdots,i_k)$.
For example, the permutation
$[\sigma(i_1),\sigma(i_2),\sigma(i_3)]$ of the ordered
list $(i_1,i_2,i_3)$ is
one of the following cases
$(i_1,i_2,i_3)$, $(i_1,i_3,i_2)$, $(i_2,i_1,i_3)$, 
$(i_2,i_3,i_1)$, $(i_3,i_1,i_2)$, or $(i_3,i_2,i_1)$.
Thus, the summation over $\sigma$
is defined, for example, by
\begin{equation}
\sum_{\sigma}
\mathcal{T}^{\sigma(i_1)\sigma(i_2)\sigma(i_3)}_{(3)}
=
\mathcal{T}^{i_1i_2i_3}_{(3)}
+\mathcal{T}^{i_1i_3i_2}_{(3)}
+\mathcal{T}^{i_2i_1i_3}_{(3)}
+\mathcal{T}^{i_2i_3i_1}_{(3)}
+\mathcal{T}^{i_3i_1i_2}_{(3)}
+\mathcal{T}^{i_3i_2i_1}_{(3)}.
\end{equation}
Summation over $\sigma$ can also be used for tensor products
like
\begin{eqnarray}
\sum_{\sigma}
\mathcal{A}^{\sigma(i_1)}_{(1)}
\mathcal{B}^{\sigma(i_2)}_{(1)}
\mathcal{C}^{\sigma(i_3)\sigma(i_4)}_{(2)}
&=&
\phantom{+}\big[\mathcal{A}^{i_1}_{(1)}\mathcal{B}^{i_2}_{(1)}
+\mathcal{A}^{i_2}_{(1)}\mathcal{B}^{i_1}_{(1)}\big]
\big[\mathcal{C}^{i_3i_4}_{(2)}+\mathcal{C}^{i_4i_3}_{(2)}\big]
\nonumber \\[-1.4ex]
&&+
\big[\mathcal{A}^{i_1}_{(1)}\mathcal{B}^{i_3}_{(1)}
+\mathcal{A}^{i_3}_{(1)}\mathcal{B}^{i_1}_{(1)}\big]
\big[\mathcal{C}^{i_2i_4}_{(2)}+\mathcal{C}^{i_4i_2}_{(2)}\big]
\nonumber \\
&&+
\big[\mathcal{A}^{i_1}_{(1)}\mathcal{B}^{i_4}_{(1)}
+\mathcal{A}^{i_4}_{(1)}\mathcal{B}^{i_1}_{(1)}\big]
\big[\mathcal{C}^{i_2i_3}_{(2)}+\mathcal{C}^{i_3i_2}_{(2)}\big]
\nonumber \\
&&+
\big[\mathcal{A}^{i_2}_{(1)}\mathcal{B}^{i_3}_{(1)}
+\mathcal{A}^{i_3}_{(1)}\mathcal{B}^{i_2}_{(1)}\big]
\big[\mathcal{C}^{i_1i_4}_{(2)}+\mathcal{C}^{i_4i_1}_{(2)}\big]
\nonumber \\
&&+
\big[\mathcal{A}^{i_2}_{(1)}\mathcal{B}^{i_4}_{(1)}
+\mathcal{A}^{i_4}_{(1)}\mathcal{B}^{i_2}_{(1)}\big]
\big[\mathcal{C}^{i_1i_3}_{(2)}+\mathcal{C}^{i_3i_1}_{(2)}\big]
\nonumber \\
&&+
\big[\mathcal{A}^{i_3}_{(1)}\mathcal{B}^{i_4}_{(1)}
+\mathcal{A}^{i_4}_{(1)}\mathcal{B}^{i_3}_{(1)}\big]
\big[\mathcal{C}^{i_1i_2}_{(2)}+\mathcal{C}^{i_2i_1}_{(2)}\big].
\end{eqnarray}
\subsection{Rotationally invariant (isotropic) tensor}
If $\mathcal{I}^{i_1\cdots i_k}_{(k)}$ is invariant under any rotation, 
\begin{equation}
\mathcal{I}^{\prime\,i_1\cdots i_k}_{(k)}=
\mathscr{R}^{\,i_1j_1}
\cdots
\mathscr{R}^{\,i_kj_k}
\mathcal{I}^{j_1\cdots j_k}_{(k)}
=
\mathcal{I}^{i_1\cdots i_k}_{(k)},
\end{equation}
then
it
is called a \textit{rotationally invariant} (\textit{isotropic})
tensor.
Because $\mathscr{R}$ is orthogonal,
$\delta^{ij}$ is the isotropic tensor
of rank 2: 
$
\mathscr{R}^{i_1j_1}
\mathscr{R}^{i_2j_2}
\delta^{j_1j_2}
=
[\mathscr{R}\mathscr{R}^T]^{i_1i_2}
=\delta^{i_1i_2}.$
In a similar manner, a product of
Kronecker deltas is always isotropic:
$
\mathscr{R}^{i_1j_1}
\mathscr{R}^{i_2j_2}
\cdots
\mathscr{R}^{i_{2k-1}j_{2k-1}}
\mathscr{R}^{i_{2k}j_{2k}}
\delta^{j_1j_2}\cdots\delta^{j_{2k-1}j_{2k}}
=
[\mathscr{R}\mathscr{R}^T]^{i_1i_2}
\cdots
[\mathscr{R}\mathscr{R}^T]^{i_{2k-1}i_{2k}}
=\delta^{i_1i_2}\cdots\delta^{i_{2k-1}i_{2k}}$. 

In general, a tensor is isotropic if and only if it is expressible 
as a linear combination of products of Kronecker deltas and possibly 
one Levi-Civita tensor: \cite{Weyl-1939,Jeffreys-1973}
\begin{equation}
\label{eq:isotropic}
\mathcal{I}^{i_1\cdots i_{k}}_{(k)}=
\sum_{\sigma}
\left[
c_{\sigma nk}
\delta^{\sigma(i_1)\sigma(i_2)}\cdots
\delta^{\sigma(i_{k-1})\sigma(i_{k})} 
+d_{\sigma nk}
\epsilon^{\sigma(i_1)\sigma(i_2)\cdots\sigma(i_n)}
\delta^{\sigma(i_{n+1})\sigma(i_{n+2})}
\cdots\delta^{\sigma(i_{k-1})\sigma(i_{k})}
\right],
\end{equation}
where 
the constants $c_{\sigma nk}$ and $d_{\sigma nk}$ depend on the
permutation $\sigma$ defined in Sec.~\ref{sec:perm}, 
the number of dimensions $n$, and the rank $k$.
The first term in the brackets survives only if $k$ is even.
The second term in the brackets survives only if $k\ge n$ and
$k-n$ is even.

\subsection{Totally symmetric isotropic tensor}
A tensor 
is called \textit{totally symmetric} 
if it is invariant
under exchange of any two vector indices.
We denote 
$\widetilde{\mathcal{I}}_{(k)}^{\,i_1\cdots i_{k}}$ by
the totally symmetric isotropic tensor of rank $k$.
Thus $\widetilde{\mathcal{I}}_{(k)}^{i_1\cdots i_{k}}$
can be constructed by symmetrizing the indices 
of terms in Eq.~(\ref{eq:isotropic}) 
with the coefficient $c_{\sigma nk}$ only.
Then non-vanishing entries are of rank even only:
\begin{equation}
\label{eq:ts-isotropic}
\widetilde{\mathcal{I}}^{i_1\cdots i_{2k}}_{(2k)}=
\frac{c_{nk}}{(2k)!!}
\sum_{\sigma}
\delta^{\sigma(i_1)\sigma(i_2)}\cdots
\delta^{\sigma(i_{2k-1})\sigma(i_{2k})},
\end{equation}
where 
$c_{nk}$ is
independent of $\sigma$ and
depends only on $n$ and $k$.
The additional factor
$(2k)!!=(2!)^k k!$ is divided to cancel the over-counts of 
the summation over $\sigma$ in Eq.~(\ref{eq:ts-isotropic}).
Here,
the factor $k!$ appears because of the commutativity 
of the multiplication of $k$ Kronecker deltas and
the factor $(2!)^k$ appears because each of $k$ Kronecker deltas
is symmetric.
As a result, the constant $c_{nk}$ is the normalization 
factor for a single \textit{distinct} product
of Kronecker deltas.
We choose the normalization 
\begin{equation}
\label{eq:normalization-i-tilde-convention}
\widetilde{\mathcal{I}}_{(2k)}^{i_1i_1i_2i_2\cdots i_{k}i_{k}}
=1,
\end{equation}
which determines the constant
$c_{nk}$ uniquely.

\subsection{Decomposition}
Any vector 
$\bm{V}\in\mathfrak{V}_{(n)}$ is expressed as the sum
of the longitudinal vector 
$\bm{V}_\parallel\in\mathfrak{V}_{\parallel(m)}$
and the transverse vector 
$\bm{V}_\perp\in\mathfrak{V}_{\perp(n-m)}$. Here,
$\mathfrak{V}_{\parallel(m)}$ and
$\mathfrak{V}_{\perp(n-m)}$ are the vector spaces spanned by
$\{\hat{\bm{e}}_p|p=1,\cdots,m\}$ defined 
in Eq.~(\ref{def:eq-from-p=1tom}) and
$\{\hat{\bm{e}}_q|\hat{\bm{e}}_q\cdot\hat{\bm{e}}_p=0
\textrm{ and }
p=1,\cdots,m\textrm{ and } q=m+1,\cdots,n\}$, respectively.
The projections onto the spaces $\mathfrak{V}_{\parallel(m)}$ and
$\mathfrak{V}_{\perp(n-m)}$ can be made by multiplying the projection
operators $\delta_\parallel^{ij}$ and $\delta_\perp^{ij}$ as
\begin{equation}
\label{eq:projection-of-vector-perp-parallel-space}
\begin{array}{ccc}
V_\parallel^i&=&\delta_\parallel^{ij}V^j,
\\
V_\perp^i&=&\delta_\perp^{ij}V^j, 
\end{array}
\end{equation}
where the longitudinal and transverse projection operators are
defined, respectively, by
\begin{eqnarray}
\label{eq:delta-parallel-perp}
\begin{array}{ccl}
\delta_{\parallel}^{ij}&=&\sum_{p=1}^m\hat{e}_p^i\hat{e}_p^j,
\\
\delta_{\perp}^{ij}&=&\sum_{q=m+1}^n\hat{e}_q^i\hat{e}_q^j,
\end{array}
\end{eqnarray}
and $\delta^{ij}=\delta_\parallel^{ij}+\delta_\perp^{ij}$.
In a similar manner, any rank-$k$
tensor $\mathcal{T}_{(k)}^{i_1\cdots i_k}$ can be decomposed as
\begin{eqnarray}
\label{eq:rank-k-vector-m}
\mathcal{T}_{(k)}^{i_1\cdots i_k}
=
(\delta_\parallel^{i_1j_1}+\delta_\perp^{i_1j_1})
\cdots
(\delta_\parallel^{i_kj_k}+\delta_\perp^{i_kj_k})
\mathcal{T}_{(k)}^{j_1\cdots j_k}.
\end{eqnarray}
For example, a rank-$k$ Cartesian tensor 
$q^{i_1}\cdots q^{i_k}$ can be decomposed as
\begin{equation}
\label{eq:decomposition-example}
q^{i_1}\cdots q^{i_k}
=
\sum_{r=0}^k
\frac{1}{r!(k-r)!}
\sum_\sigma
q_\parallel^{\sigma(i_1)}
\cdots
q_\parallel^{\sigma(i_r)}
q_\perp^{\sigma(i_{r+1})}
\cdots 
q_\perp^{\sigma(i_k)},
\end{equation}
where the factor $r!(k-r)!$ is divided for a given $r$
to cancel the over-counted permutations of $r$ $q_\parallel$'s
and $k-r$ $q_\perp$'s that are identical, respectively.
For the rank-3 case $q^{i}q^{j}q^{k}$,
the explicit expansion is given by
\begin{equation}
q^{i}q^jq^k
=
q_\perp^i q_\perp^j q_\perp^k
+(q_\parallel^i q_\perp^j q_\perp^k
+q_\perp^i q_\parallel^j q_\perp^k
+q_\perp^i q_\perp^j q_\parallel^k)
+
(
q_\perp^i q_\parallel^j q_\parallel^k
+
q_\parallel^i q_\perp^j q_\parallel^k
+
q_\parallel^i q_\parallel^j q_\perp^k
)
+q_\parallel^i q_\parallel^j q_\parallel^k.
\end{equation}

\section{Totally symmetric isotropic tensor
$\bm{\widetilde{\mathcal{I}}^{i_1\cdots i_{2k}}_{(2k)}}$
\label{sec:ans}}
In this section,
we carry out a rigorous derivation
of the totally symmetric isotropic tensors
$\widetilde{\mathcal{I}}^{i_1\cdots i_{2k}}_{(2k)}$
of arbitrary ranks in $\mathfrak{V}_{(n)}$
based on only the symmetries of the abstract algebraic structure.

\subsection{Recurrence relation}
According to Eq.~(\ref{eq:ts-isotropic}), 
$\widetilde{\mathcal{I}}_{(2k)}^{i_1i_2\cdots i_{2k}}$ 
must be a linear combination of products of 
Kronecker deltas.
If we factor out $\delta^{i_1j}$ for $j=i_2,\cdots,i_{2k}$,
then its coefficient must be a
totally symmetric
isotropic
tensor of rank $2k-2$
so that
\begin{equation}
\label{T-2k-A-T-series}
\widetilde{\mathcal{I}}_{(2k)}^{i_1i_2\cdots i_{2k}}
=
\lambda_{2k}
\left[
\delta^{i_1i_2}
\widetilde{\mathcal{I}}_{(2k-2)}^{i_3\cdots i_{2k}}+
\delta^{i_1i_3}
\widetilde{\mathcal{I}}_{(2k-2)}^{i_2i_4\cdots i_{2k}}+\cdots+
\delta^{i_1i_{2k}}
\widetilde{\mathcal{I}}_{(2k-2)}^{i_2\cdots i_{2k-1}}
\right],
\end{equation}
where $\lambda_{2k}$ is a constant and
there are $2k-1$ terms in the brackets.
If we multiply 
$\delta^{i_1i_2}
\delta^{i_3i_4}
\cdots
\delta^{i_{2k-1}i_{2k}}$ and impose 
the normalization condition 
in Eq.~(\ref{eq:normalization-i-tilde-convention}),
then the term in the brackets proportional to 
$\delta^{i_1i_2}$ gives $n$
and each of the remaining $2k-2$ terms gives unity.
As a result, we determine
$\lambda_{2k}=1/(n+2k-2)$ and the recurrence 
relation as
\begin{equation}
\label{T-2k-A-T-series-norm}
\widetilde{\mathcal{I}}_{(2k)}^{i_1i_2\cdots i_{2k}}
=
\frac{1}{n+2k-2}
\big[
\delta^{i_1i_2}\widetilde{\mathcal{I}}_{(2k-2)}^{i_3\cdots i_{2k}}+
\delta^{i_1i_3}\widetilde{\mathcal{I}}_{(2k-2)}^{i_2i_4\cdots i_{2k}}+\cdots+
\delta^{i_1i_{2k}}\widetilde{\mathcal{I}}_{(2k-2)}^{i_2\cdots i_{2k-1}}
\big].
\end{equation}

\subsection{Complete reduction into Kronecker deltas
\label{subsec:induction}}
We are ready to find the explicit form of 
$\widetilde{\mathcal{I}}_{(2k)}^{i_1i_2\cdots i_{2k}}$
by making recursive use of 
Eq.~(\ref{T-2k-A-T-series-norm}).
Substituting $k=1$ 
into Eq.~(\ref{T-2k-A-T-series-norm}), we determine 
$\widetilde{\mathcal{I}}_{(2)}^{ij}$ as
\begin{equation}
\label{eq:ak-final-1}
\widetilde{\mathcal{I}}_{(2)}^{ij}=\frac{\delta^{ij}}{n},
\end{equation}
where we have set $\widetilde{\mathcal{I}}_{(0)}=1$
to be consistent with the normalization 
in Eq.~(\ref{eq:normalization-i-tilde-convention}).
In this manner, we can find
$\widetilde{\mathcal{I}}_{(2k+2)}^{i_1\cdots i_{2k+2}}$
once $\widetilde{\mathcal{I}}_{(2k)}^{i_1i_2\cdots i_{2k}}$
is known. The next two entries are given by
\begin{subequations}
\label{T2K-final-result}
\begin{eqnarray}
\label{T4-final-result}
\widetilde{\mathcal{I}}_{(4)}^{i_1i_2i_3i_4}&=&
\frac{1}{n+2}
\big[
\delta^{i_1i_2}\widetilde{\mathcal{I}}_{(2)}^{i_3i_4}
+\delta^{i_1i_3}\widetilde{\mathcal{I}}_{(2)}^{i_2i_4}
+\delta^{i_1i_4}\widetilde{\mathcal{I}}_{(2)}^{i_2i_3}
\big]
\nonumber\\
&=&\frac{1}{n(n+2)}
\big[\delta^{i_1i_2}\delta^{i_3i_4}
+\delta^{i_1i_3}\delta^{i_2i_4}
+\delta^{i_1i_4}\delta^{i_2i_3}\big],
\\
\label{T6-final-result}
\widetilde{\mathcal{I}}_{(6)}^{i_1i_2i_3i_4i_5i_6}&=&
\frac{1}{n+4}
\big[
 \delta^{i_1i_2}\widetilde{\mathcal{I}}_{(4)}^{i_3i_4i_5i_6}
+\delta^{i_1i_3}\widetilde{\mathcal{I}}_{(4)}^{i_2i_4i_5i_6}
+\delta^{i_1i_4}\widetilde{\mathcal{I}}_{(4)}^{i_2i_3i_5i_6}
+\delta^{i_1i_5}\widetilde{\mathcal{I}}_{(4)}^{i_2i_3i_4i_6}
+\delta^{i_1i_6}\widetilde{\mathcal{I}}_{(4)}^{i_2i_3i_4i_5}
\big]
\nonumber\\
&=&
\frac{1}{n(n+2)(n+4)}
\big[\,\,
 \delta^{i_1i_2}
 \big(
 \delta^{i_3i_4}\delta^{i_5i_6}
+\delta^{i_3i_5}\delta^{i_4i_6}
+\delta^{i_3i_6}\delta^{i_4i_5}
 \big)
\nonumber\\
&& 
\quad\quad\quad\quad\quad\quad\quad
+\delta^{i_1i_3}
 \big(
 \delta^{i_2i_4}\delta^{i_5i_6}
+\delta^{i_2i_5}\delta^{i_4i_6}
+\delta^{i_2i_6}\delta^{i_4i_5}
 \big)
 \nonumber\\
&& 
\quad\quad\quad\quad\quad\quad\quad
+\delta^{i_1i_4}
 \big(
 \delta^{i_3i_2}\delta^{i_5i_6}
+\delta^{i_3i_5}\delta^{i_2i_6}
+\delta^{i_3i_6}\delta^{i_2i_5}
 \big)
 \nonumber\\
&& 
\quad\quad\quad\quad\quad\quad\quad
+\delta^{i_1i_5}
 \big(
 \delta^{i_3i_2}\delta^{i_5i_6}
+\delta^{i_3i_5}\delta^{i_2i_6}
+\delta^{i_3i_6}\delta^{i_2i_5}
 \big)
 \nonumber\\
&& 
\quad\quad\quad\quad\quad\quad\quad
+\delta^{i_1i_6}
 \big(
 \delta^{i_3i_4}\delta^{i_5i_2}
+\delta^{i_3i_5}\delta^{i_4i_2}
+\delta^{i_3i_2}\delta^{i_4i_5}
 \big)
 \big].
\end{eqnarray}
\end{subequations}
Each term in the brackets of the second equalities in  
Eq.~(\ref{T2K-final-result}) represents 
a single distinct product of Kronecker deltas.
In general, 
the normalization factor
$c_{nk}$ in Eq.~(\ref{eq:ts-isotropic}) is determined as
\begin{equation}
c_{nk}=\frac{1}{n(n+2)\cdots (n+2k-2)}.
\end{equation}
Our final result for the explicit
form of $\widetilde{\mathcal{I}}_{(2k)}^{i_1i_2\cdots i_{2k-1}i_{2k}}$
is
\begin{equation}
\label{eq:N-2k-summation}
\widetilde{\mathcal{I}}_{(2k)}^{i_1i_2\cdots i_{2k-1}i_{2k}}
=
\frac{1}{n(n+2)\cdots (n+2k-2)}
\frac{1}{(2k)!!}
\sum_{\sigma}
\delta^{\sigma(i_1)\sigma(i_2)}\cdots
\delta^{\sigma(i_{2k-1})\sigma(i_{2k})}.
\end{equation}
While the summation in Eq.~(\ref{eq:N-2k-summation})
is over the $(2k)!$ permutations of $(i_1,\cdots,i_{2k})$,
the number of distinct terms in 
$\widetilde{\mathcal{I}}_{(2k)}^{i_1i_2\cdots i_{2k-1}i_{2k}}$
is $\mathcal{N}_{(2k)}\equiv(2k)!/(2k)!!=(2k-1)!!$.

\subsection{Projection operator}
The normalization condition 
$\widetilde{\mathcal{I}}_{(2k)}^{i_1i_1i_2i_2\cdots i_{k}i_{k}}
=1$
in Eq.~(\ref{eq:normalization-i-tilde-convention})
requires that
\begin{eqnarray}
\widetilde{\mathcal{I}}_{(2k)}^{i_1\cdots i_{2k}}
\widetilde{\mathcal{I}}_{(2k)}^{i_1\cdots i_{2k}}
&=&\mathcal{N}_{(2k)}c_{nk}
\delta^{i_1i_2}\cdots \delta^{i_{2k-1}i_{2k}}
\widetilde{\mathcal{I}}_{(2k)}^{i_1\cdots i_{2k}}
\nonumber \\
&=&\mathcal{N}_{(2k)} c_{nk}
\nonumber \\
&=&
\frac{(2k-1)!!}{n(n+2)\cdots(n+2k-2)},
\end{eqnarray}
where we have made use of the
symmetric property of 
$\widetilde{\mathcal{I}}_{(2k)}^{i_1\cdots i_{2k}}$:
Every one of $\mathcal{N}_{(2k)}$ distinct terms in 
a factor
$\widetilde{\mathcal{I}}_{(2k)}^{i_1\cdots i_{2k}}$ 
on the left side
has the identical contribution
$c_{nk}\delta^{i_1i_2}\cdots \delta^{i_{2k-1}i_{2k}}$
to the product.
It is straightforward to project out the totally symmetric
isotropic part $\widetilde{\mathcal{T}}^{i_1\cdots i_{2k}}_{(2k)}$ of a tensor $\mathcal{T}_{(2k)}^{i_1\cdots i_{2k}}$ as
\begin{subequations}
\label{eq:projection-of-t-to-totally-sym}
\begin{equation}
\widetilde{\mathcal{T}}^{i_1\cdots i_{2k}}_{(2k)}
=
\Pi^{i_1\cdots i_{2k};j_1\cdots j_{2k}}
\mathcal{T}^{j_1\cdots j_{2k}}_{(2k)},
\end{equation}
where the projection operator 
$\Pi^{i_1\cdots i_{2k};j_1\cdots j_{2k}}$ is
defined by 
\begin{equation}
\Pi^{i_1\cdots i_{2k};j_1\cdots j_{2k}}
\equiv
\frac{n(n+2)\cdots(n+2k-2)}{(2k-1)!!}
\,
\widetilde{\mathcal{I}}_{(2k)}^{i_1\cdots i_{2k}}
\widetilde{\mathcal{I}}_{(2k)}^{j_1\cdots j_{2k}}.
\end{equation}
\end{subequations}

\section{Application to Tensor Integrals\label{sec:app}}
In this section, we apply the result in Eq.~(\ref{eq:N-2k-summation})
for the totally symmetric isotropic tensor of arbitrary
rank in $n$ dimensions
to compute various tensor integrals involving angle averages.
This leads to a great simplification 
of the evaluation steps into simple
counts of combinatorics.

\subsection{Angle average}
Let us consider the tensor integral
\begin{equation}
\label{isotropic-angular-integral}
\langle \hat{r}^{i_1}
\cdots
\hat{r}^{i_k}
\rangle_{\hat{\bm{r}}}
\equiv
\frac{1}{\Omega^{(n)}}
\int d\Omega_{\hat{\bm{r}}}^{(n)}\,
\hat{r}^{i_1}
\cdots
\hat{r}^{i_k},
\end{equation}
where $\hat{\bm{r}}\equiv(\hat{r}^{1},
\hat{r}^{2},\cdots
\hat{r}^{n})$ is the unit radial vector $(\hat{\bm{r}}^2=1)$
and the tensor is independent of any specific vectors.
The analytic expressions for the $n$-dimensional solid angle 
$\Omega^{(n)}$ and its differential element 
$d\Omega^{(n)}_{\hat{\bm{r}}}$ expressed in terms of 
$n-2$ polar angles and an azimuthal angle are listed in
Appendix~\ref{app:polar-azimuthal}.

It is manifest that 
$
\langle \hat{r}^{i_1}
\cdots
\hat{r}^{i_k}
\rangle_{\hat{\bm{r}}}
$
is isotropic and totally symmetric
and, therefore, it must be proportional to 
$\widetilde{\mathcal{I}}_{(k)}^{i_1\cdots i_{k}}$.
Because $\widetilde{\mathcal{I}}_{(k)}^{i_1\cdots i_{k}}=0$
for any $k$ odd, the only non-vanishing components are
$
\langle \hat{r}^{i_1}
\cdots
\hat{r}^{i_{2k}}
\rangle_{\hat{\bm{r}}}
$.
By multiplying 
$\delta^{i_1i_2}\delta^{i_3i_4}\cdots \delta^{i_{2k-1}i_{2k}}$, 
summing over the indices, and 
substituting $\hat{\bm{r}}^2=1$,
we find that $\langle \hat{r}^{i_1}
\cdots
\hat{r}^{i_{2k}}
\rangle_{\hat{\bm{r}}}$
satisfies the normalization condition 
in Eq.~(\ref{eq:normalization-i-tilde-convention}). 
As a result,
\begin{equation}
\label{eq:angle-average-i-tilde}
\langle \hat{r}^{i_1}
\cdots
\hat{r}^{i_{2k}}
\rangle_{\hat{\bm{r}}}
=
\widetilde{\mathcal{I}}_{(2k)}^{i_1\cdots i_{2k}}.
\end{equation}
This
can be applied to derive the angle-average formulas 
$\langle
(\bm{a}\cdot\hat{\bm{r}})^{2k+1}\rangle_{\hat{\bm{r}}}
=0$
and
\begin{equation}
\label{eq:ar-2k-wo-angle-result}
\langle
(\bm{a}\cdot\hat{\bm{r}})^{2k}\rangle_{\hat{\bm{r}}}
=
a^{i_1}a^{i_2}\cdots a^{i_{2k}}\,\widetilde{\mathcal{I}}_{(2k)}^{i_1i_2\cdots i_{2k}}
=\frac{(2k-1)!!}{n(n+2)\cdots (n+2k-2)}\bm{a}^{2k},
\end{equation}
where $\bm{a}$ is a constant vector
and $k$ is a positive integer.
This result agrees with  
Eq.~(\ref{eq:ar-2k-angle-result}) that is obtained by 
direct evaluations of angular integrals.
In general,
for any constant vectors $\bm{a}_1,\cdots, \bm{a}_m$, we obtain
\begin{equation}
\label{eq:ar-general-result}
\langle
(\bm{a_1}\cdot\hat{\bm{r}})
(\bm{a_2}\cdot\hat{\bm{r}})
\cdots
(\bm{a_m}\cdot\hat{\bm{r}})
\rangle_{\hat{\bm{r}}}
=
a_1^{i_1}a_2^{i_2}\cdots a_m^{i_{m}}
\,\widetilde{\mathcal{I}}_{(m)}^{i_1i_2\cdots i_{m}}.
\end{equation}
The expression vanishes for all $m$ odd and 
$\bm{a}_i$'s do not have to be distinct.
As is discussed in Appendix \ref{app:explicit-evaluation-adotkton},
an explicit integration over polar and 
azimuthal angles is
 extremely tedious even in two or three dimensions
and it is nontrivial to obtain the general form
in Eq.~(\ref{eq:ar-general-result}) directly by integration.
Therefore, our strategy to make use of
$\widetilde{\mathcal{I}}_{(m)}^{i_1i_2\cdots i_{m}}$
is a quite efficient way to evaluate the angular 
integrals.

\subsection{Tensor-integral reduction}
In general, the integrand of a tensor angular integral
may have a scalar factor that depends on constant vectors
as well as the integral variable $\bm{q}$,
while the angle average in Eq.~(\ref{isotropic-angular-integral})
is independent of any specific vectors.
The simplest case is that the scalar factor depends only on 
$\bm{q}$:
\begin{equation}
\label{tensor-A}
\mathcal{A}^{i_1\cdots i_k}_{(k)}
=
\int_{\bm{q}}
q^{i_1}\cdots q^{i_k}f(\bm{q}),
\end{equation}
where $\bm{q}=(q^1,\cdots,q^n)$,
$f(\bm{q})$ is a scalar, and
the symbol $\int_{\bm{q}}$ is defined by
\begin{equation}
\int_{\bm{q}}
\equiv
\int_{-\infty}^\infty dq^1 \cdots
\int_{-\infty}^\infty dq^n 
=
\int_0^\infty d|\bm{q}|\,\, |\bm{q}|^{n-1}
\int d\Omega^{(n)}_{\hat{\bm{q}}}.
\end{equation}
Here,
$d\Omega^{(n)}_{\hat{\bm{q}}}$ is the
differential solid-angle element of $\hat{\bm{q}}$.
It is manifest that $\mathcal{A}^{i_1\cdots i_k}_{(k)}$
in Eq.~(\ref{tensor-A}) is totally symmetric and isotropic
so that only the even-rank case survives:
$\mathcal{A}^{i_1\cdots i_{2k}}_{(2k)}
=\Pi^{i_1\cdots i_{2k};j_1\cdots j_{2k}}
\mathcal{A}^{j_1\cdots j_{2k}}_{(2k)}$
according to Eq.~(\ref{eq:projection-of-t-to-totally-sym}).
Therefore,
the non-vanishing elements are completely
determined as
\begin{equation}
\label{eq:tensor-A-reduced}
\mathcal{A}^{i_1\cdots i_{2k}}_{(2k)}
=\widetilde{\mathcal{I}}_{(2k)}^{i_1\cdots i_{2k}}
\int_{\bm{q}}\bm{q}^{2k}
f(\bm{q}),
\end{equation}
where we have used
$\widetilde{\mathcal{I}}_{(2k+1)}^{i_1\cdots i_{2k+1}}
q^{i_1}\cdots q^{i_{2k+1}}=0$ and
$\widetilde{\mathcal{I}}_{(2k)}^{i_1\cdots i_{2k}}
q^{i_1}\cdots q^{i_{2k}}=
\frac{(2k-1)!!}{n(n+2)\cdots (n+2k-2)}\bm{q}^{2k}$.
As a result, 
one has to compute only a single scalar
integral $\int_{\bm{q}}\bm{q}^{2k}f(\bm{q})$
without evaluating all of the $n^{2k}$ components
of $\mathcal{A}^{i_1\cdots i_{2k}}_{(2k)}$.

A more complicated situation is that
\begin{equation}
\label{tensor-B}
\mathcal{B}^{i_1\cdots i_k}_{(k)}
=
\int_{\bm{q}}
q^{i_1}\cdots q^{i_k}g(\bm{q},\bm{a}),
\end{equation}
where $g(\bm{q},\bm{a})$ is a scalar that
depends on $\bm{q}$ and a constant vector $\bm{a}$.
It is manifest that $\mathcal{B}^{i_1\cdots i_k}_{(k)}$ 
is symmetric under 
exchange of any two vector indices and
it must depend only on $\bm{a}$ because
$\bm{q}$ is integrated out.
We call $\mathfrak{V}_{\parallel (1)}$
the one-dimensional Euclidean space spanned by $\bm{a}$
and $\mathfrak{V}_{\perp(n-1)}$ the space perpendicular to $\bm{a}$.
According to Eq.~(\ref{eq:projection-of-vector-perp-parallel-space}),
$q^i=q_\parallel^i+q_\perp^i$ with 
$\bm{q}_\parallel
=(\bm{q}\cdot\hat{\bm{a}})\hat{\bm{a}}\in \mathfrak{V}_{\parallel(1)}$
and
$\bm{q}_\perp=\bm{q}-\bm{q}_\parallel\in \mathfrak{V}_{\perp(n-1)}$.
By making use of Eq.~(\ref{eq:decomposition-example}), 
we decompose $\mathcal{B}^{i_1\cdots i_k}_{(k)}$ as
\begin{equation}
\label{tensor-B-3}
\mathcal{B}^{i_1\cdots i_k}_{(k)}
=
\sum_{r=0}^k
\frac{1}{r!(k-r)!}
\sum_{\sigma}
\,\,\hat{a}^{\sigma(i_1)}\cdots\hat{a}^{\sigma(i_r)}\!
\int_{\bm{q}}
q_\perp^{\sigma(i_{r+1})}\cdots q_\perp^{\sigma(i_k)}g_r(\bm{q},\bm{a}),
\end{equation}
where 
$g_r(\bm{q},\bm{a})\equiv(\bm{q}\cdot\hat{\bm{a}})^rg(\bm{q},\bm{a})$.
The first summation is over the number of longitudinal components, $r$.
The constant tensor $\hat{a}^{\sigma(i_1)}\cdots\hat{a}^{\sigma(i_r)}$
of rank $r$ and
the remaining tensor integral 
$\int_{\bm{q}}
q_\perp^{\sigma(i_{r+1})}\cdots q_\perp^{\sigma(i_k)}g_r(\bm{q},\bm{a})$
of rank $k-r$
are defined in $\mathfrak{V}_{\parallel(1)}$
and $\mathfrak{V}_{\perp(n-1)}$, respectively.
The tensor integral in $\mathfrak{V}_{\perp(n-1)}$
is totally symmetric and isotropic. Thus
we find that
\begin{equation}
\int_{\bm{q}}
q_\perp^{\sigma(i_{r+1})}\cdots q_\perp^{\sigma(i_k)}g_r(\bm{q},\bm{a})
=\widetilde{\mathcal{I}}_{\perp(k-r)}
^{\sigma(i_{r+1})\cdots \sigma(i_{k})}
\int_{\bm{q}}|\bm{q}_{\perp}|^{k-r}
g_r(\bm{q},\bm{a}),
\end{equation}
where $\widetilde{\mathcal{I}}_{\perp(k)}^{i_{1}\cdots i_{k}}$
is the $\mathfrak{V}_{\perp(n-1)}$ analogue of 
$\widetilde{\mathcal{I}}_{(k)}^{i_{1}\cdots i_{k}}$ defined 
in $\mathfrak{V}_{(n)}$:
The explicit form of 
$\widetilde{\mathcal{I}}_{\perp(k)}^{i_{1}\cdots i_{k}}$ 
can be obtained by replacing every Kronecker delta
with the corresponding one in $\mathfrak{V}_{\perp(n-1)}$
that is given in Eq.~(\ref{eq:delta-parallel-perp})
and replacing the dimension $n$ with $n-1$ as
\begin{equation}
\widetilde{\mathcal{I}}_{\perp(k)}^{i_{1}\cdots i_{k}}
=
\widetilde{\mathcal{I}}_{(k)}^{i_{1}\cdots i_{k}}
\Big|_{\delta^{ij}\to\delta_\perp^{ij},\,n\to n-1}.
\end{equation}
As a result, $\mathcal{B}^{i_1\cdots i_{k}}_{(k)}$
can be expressed as
the following linear combination:
\begin{equation}
\label{tensor-B-even}
\mathcal{B}^{i_1\cdots i_{k}}_{(k)}
=
\sum_{r=0}^k
{\mathcal{E}}_{(r|k)}^{i_1\cdots i_{k}}
\int_{\bm{q}}
\,|\bm{q}_\perp|^{k-r}g_r(\bm{q},\bm{a}),
\end{equation}
where the constant tensor 
${\mathcal{E}}_{(r|k)}^{i_1\cdots i_{k}}$
of rank $k$ with $r$ longitudinal indices is defined by
\begin{equation}
\mathcal{E}_{(r|k)}^{i_1\cdots i_k}
=
\begin{cases}
\displaystyle
\frac{1}{r!(k-r)!}
\sum_{\sigma}
\hat{a}^{\sigma(i_{1})}
\cdots
\hat{a}^{\sigma(i_{r})}\,
\widetilde{\mathcal{I}}_{\perp(k-r)}^{\sigma(i_{r+1})\cdots \sigma(i_{k})},
& \textrm{if $k-r$ is even},
\\
0,& \textrm{if $k-r$ is odd}.
\end{cases}
\end{equation}
Here, $\mathcal{E}_{(r|k)}^{i_1\cdots i_k}$ is 
totally symmetric. Although this tensor is not
completely isotropic, there is a partial isotropy among
transverse components only.
For example, we list the reduction formulas for
the first four entries of 
$\mathcal{B}^{i_1\cdots i_{k}}_{(k)}$:
\begin{eqnarray}
\int_{\bm{q}}
{q}^i
g(\bm{q},\bm{a})&=&
\mathcal{E}_{(1|1)}^{i}
\int_{\bm{q}}
\,
g_1(\bm{q},\bm{a}),
\nonumber
\\
\int_{\bm{q}}
{q}^i
{q}^j
g(\bm{q},\bm{a})
&=&
\mathcal{E}_{(0|2)}^{ij}
\int_{\bm{q}}
\bm{q}_\perp^2
g_0(\bm{q},\bm{a})
+
\mathcal{E}_{(2|2)}^{ij}
\int_{\bm{q}}
g_2(\bm{q},\bm{a}),
\nonumber\\
\int_{\bm{q}}
{q}^i
{q}^j
{q}^k
g(\bm{q},\bm{a})
&=&
\mathcal{E}_{(1|3)}^{ijk}
\int_{\bm{q}}
\bm{q}_\perp^2
g_1(\bm{q},\bm{a})
+
\mathcal{E}_{(3|3)}^{ijk}
\int_{\bm{q}}
g_3(\bm{q},\bm{a}),
\nonumber\\
\int_{\bm{q}}
{q}^i
{q}^j
{q}^k
{q}^\ell
g(\bm{q},\bm{a})
&=&
\mathcal{E}_{(0|4)}^{ijk\ell}
\int_{\bm{q}}
\bm{q}_\perp^4
g_0(\bm{q},\bm{a})
+\mathcal{E}_{(2|4)}^{ijk\ell}
\int_{\bm{q}}
\bm{q}_\perp^2
g_2(\bm{q},\bm{a})
+\mathcal{E}_{(4|4)}^{ijk\ell}
\int_{\bm{q}}
g_4(\bm{q},\bm{a}),
\end{eqnarray}
where
\begin{eqnarray}
\mathcal{E}_{(1|1)}^{i}&=&\hat{a}^i,
\nonumber \\
\mathcal{E}_{(0|2)}^{ij}
&=&
\frac{\delta_\perp^{ij}}{n-1},
\nonumber \\
\mathcal{E}_{(2|2)}^{ij}
&=&
\hat{a}^i\hat{a}^j,
\nonumber \\
\mathcal{E}_{(1|3)}^{ijk}
&=&
\frac{1}{n-1}
\big(
\hat{a}^i\delta_\perp^{jk}+
\hat{a}^j\delta_\perp^{ki}+
\hat{a}^k\delta_\perp^{ij}
\big),
\nonumber \\
\mathcal{E}_{(3|3)}^{ijk}
&=&
\hat{a}^i\hat{a}^j\hat{a}^k,
\nonumber \\
\mathcal{E}^{ijk\ell}_{(0|4)}
&=&\frac{1}{(n-1)(n+1)}
\big(
\delta_\perp^{ij}\delta_\perp^{k\ell}+
\delta_\perp^{ik}\delta_\perp^{j\ell}+
\delta_\perp^{i\ell}\delta_\perp^{jk}
\big),
\nonumber \\
\mathcal{E}_{(2|4)}^{ijk\ell}
&=&
\frac{1}{n-1}
\big(
\hat{a}^i\hat{a}^j\delta_\perp^{k\ell}+
\hat{a}^i\hat{a}^k\delta_\perp^{j\ell}+
\hat{a}^i\hat{a}^\ell\delta_\perp^{kj}+
\hat{a}^j\hat{a}^k\delta_\perp^{\ell i}+
\hat{a}^j\hat{a}^\ell\delta_\perp^{k i}+
\hat{a}^k\hat{a}^\ell\delta_\perp^{ij}
\big),
\nonumber \\
\mathcal{E}_{(4|4)}^{ijk\ell}
&=&
\hat{a}^i\hat{a}^j\hat{a}^k\hat{a}^\ell.
\end{eqnarray}

The most general case is that 
the integrand
depends on $m$ linearly independent
constant vectors $\bm{a}_1$, $\cdots$, $\bm{a}_m$:
\begin{equation}
\label{tensor-C}
\mathcal{C}^{i_1\cdots i_k}_{(k)}
=
\int_{\bm{q}}
q^{i_1}\cdots q^{i_k}h(\bm{q},\bm{a}_1,\cdots,\bm{a}_m),
\end{equation}
where $h(\bm{q},\bm{a}_1,\cdots,\bm{a}_m)$ is a scalar.
The tensor $\mathcal{C}^{i_1\cdots i_k}_{(k)}$ is
symmetric under exchange of any two vector indices.
We call $\mathfrak{V}_{\parallel (m)}$
the $m$-dimensional Euclidean space spanned by 
$\bm{a}_1,\cdots,\bm{a}_m$,
and $\mathfrak{V}_{\perp(n-m)}$ the space perpendicular to
those constant vectors.
We choose the unit basis vectors $\hat{\bm{e}}_1,\cdots,\hat{\bm{e}}_m$
in Eq.~(\ref{def:eq-from-p=1tom})
to span $\mathfrak{V}_{\parallel (m)}$.
Then we can decompose $\bm{q}$ as
$\bm{q}=\bm{q}_\parallel+\bm{q}_\perp$, where
$\bm{q}_\parallel=\sum_{p=1}^m
\hat{\bm{e}}_p(\hat{\bm{e}}_p\cdot\bm{q})\in 
\mathfrak{V}_{\parallel (m)}$ and 
$\bm{q}_\perp=\bm{q}-\bm{q}_\parallel\in \mathfrak{V}_{\perp(n-m)}$.
By making use of the identity in Eq.~(\ref{eq:decomposition-example}),
we can express
$\mathcal{C}^{i_1\cdots i_k}_{(k)}$ as
\begin{equation}
\label{tensor-C-2}
\mathcal{C}^{i_1\cdots i_k}_{(k)}
=
\sum_{r=0}^k
\frac{1}{r!(k-r)!}
\sum_{\sigma}
\int_{\bm{q}}
{q}_\parallel^{\sigma(i_1)}
\cdots
{q}_\parallel^{\sigma(i_r)}
q_\perp^{\sigma(i_{r+1})}\cdots q_\perp^{\sigma(i_k)}h(\bm{q},\bm{a}_1,\cdots,\bm{a}_m),
\end{equation}
where the first summation is over $r$, the number 
of longitudinal components.
Because every $\hat{{e}}^i_p$ is independent of $\bm{q}$,
we find that 
\begin{eqnarray}
\label{tensor-C-3}
\mathcal{C}^{i_1\cdots i_k}_{(k)}
&=&
\sum_{r=0}^k
\frac{1}{r!(k-r)!}
\sum_{p_1,\cdots,\,p_r=1}^{m}
\sum_{\sigma}
\hat{e}_{p_1}^{\sigma(i_1)}
\cdots
\hat{e}_{p_r}^{\sigma(i_r)}
\,
\widetilde{\mathcal{I}}_{\perp(k-r)}^{\sigma(i_{r+1})\cdots \sigma(i_k)}
\nonumber \\
&&\times
\int_{\bm{q}}
|\bm{q}_\perp|^{k-r}
h_{p_1,\cdots,\,p_r}
(\bm{q},\bm{a}_1,\cdots,\bm{a}_m),
\end{eqnarray}
where $p_1,\cdots, p_r$ are indices of Cartesian axes
of $\mathfrak{V}_{\parallel (m)}$ and
the scalar $h_{p_1,\cdots, \,p_r}$ is defined by
\begin{equation}
h_{p_1,\cdots, \,p_r}(\bm{q},\bm{a}_1,\cdots,\bm{a}_m)
\equiv 
(\bm{q}\cdot \hat{\bm{e}}_{p_1})\cdots
(\bm{q}\cdot \hat{\bm{e}}_{p_r})
\,h(\bm{q},\bm{a}_1,\cdots,\bm{a}_m).
\end{equation}
In Eq.~(\ref{tensor-C-3}), 
we have made a replacement 
$q_\perp^{\sigma(i_{r+1})}\cdots q_\perp^{\sigma(i_k)}
\to
\widetilde{\mathcal{I}}_{\perp(k-r)}^{\sigma(i_{r+1})\cdots \sigma(i_k)}
|\bm{q}_\perp|^{k-r}$ by taking into account
the isotropy of $\mathcal{C}^{i_1\cdots i_k}_{(k)}$
in the space $\mathfrak{V}_{\perp(n-m)}$.
Here, $\widetilde{\mathcal{I}}_{\perp(k)}^{i_{1}\cdots i_{k}}$
is the $\mathfrak{V}_{\perp(n-m)}$ analogue of 
$\widetilde{\mathcal{I}}_{(k)}^{i_{1}\cdots i_{k}}$ 
that is defined 
in $\mathfrak{V}_{(n)}$:
The explicit form of 
$\widetilde{\mathcal{I}}_{\perp(k)}^{i_{1}\cdots i_{k}}$ 
can be obtained by replacing every Kronecker delta
with the corresponding one in $\mathfrak{V}_{\perp(n-m)}$
that is given in Eq.~(\ref{eq:delta-parallel-perp})
and replacing the dimension $n$ with $n-m$ as
\begin{equation}
\widetilde{\mathcal{I}}_{\perp(k)}^{i_{1}\cdots i_{k}}
=
\widetilde{\mathcal{I}}_{(k)}^{i_{1}\cdots i_{k}}
\Big|_{\delta^{ij}\to\delta_\perp^{ij},\,n\to n-m}.
\end{equation}
As an example, we list
the first four entries of
 the tensor-integral reduction formulas
for $m=2$:
\begin{eqnarray}
\int_{\bm{q}}q^i\,h(\bm{q},\bm{a}_1,\bm{a}_2)
&=&
\sum_{a=1}^2
\hat{e}_a^i
\int_{\bm{q}}\,h_a(\bm{q},\bm{a}_1,\bm{a}_2),
\nonumber
\\
\int_{\bm{q}}q^iq^j\,h(\bm{q},\bm{a}_1,\bm{a}_2)
&=&
\sum_{a,b=1}^2
\hat{e}_a^i
\hat{e}_b^j
\int_{\bm{q}}
\,h_{a,b}(\bm{q},\bm{a}_1,\bm{a}_2)
+\frac{\delta_\perp^{ij}}{n-2}
\int_{\bm{q}}\bm{q}_\perp^2
\,h(\bm{q},\bm{a}_1,\bm{a}_2),
\nonumber
\\
\int_{\bm{q}}q^iq^jq^k\,h(\bm{q},\bm{a}_1,\bm{a}_2)
&=&
\sum_{a,b,c=1}^2
\hat{e}_a^i\hat{e}_b^j\hat{e}_c^k
\int_{\bm{q}}
\,h_{a,b,c}(\bm{q},\bm{a}_1,\bm{a}_2)
\nonumber\\
&&
+
\sum_{a=1}^2
\frac{
\hat{e}_a^i\delta_\perp^{jk}+
\hat{e}_a^j\delta_\perp^{ki}+
\hat{e}_a^k\delta_\perp^{ij}}{n-2}
\int_{\bm{q}}
\bm{q}_\perp^2
\,h_a(\bm{q},\bm{a}_1,\bm{a}_2),
\nonumber 
\end{eqnarray}
\begin{eqnarray}
\int_{\bm{q}}q^iq^jq^kq^\ell\,h(\bm{q},\bm{a}_1,\bm{a}_2)
&=&
\sum_{a,b,c,d=1}^2
\hat{e}_a^i\hat{e}_b^j\hat{e}_c^k\hat{e}_d^\ell
\int_{\bm{q}}
\,h_{a,b,c,d}(\bm{q},\bm{a}_1,\bm{a}_2)
\nonumber\\
&&
+
\sum_{a,b=1}^2
\frac{\hat{e}_a^i\hat{e}_b^j\delta_\perp^{k\ell}+
\hat{e}_a^i\hat{e}_b^k\delta_\perp^{j\ell}+
\hat{e}_a^i\hat{e}_b^\ell\delta_\perp^{jk}+
\hat{e}_a^k\hat{e}_b^\ell\delta_\perp^{ij}+
\hat{e}_a^j\hat{e}_b^\ell\delta_\perp^{ik}+
\hat{e}_a^j\hat{e}_b^k\delta_\perp^{i\ell}}{n-2}
\nonumber\\&&
\quad\quad\times
\int_{\bm{q}}
\bm{q}_\perp^2
\,h_{a,b}(\bm{q},\bm{a}_1,\bm{a}_2)
\nonumber\\
&&
+
\frac{\delta_\perp^{ij}\delta_\perp^{k\ell}+
\delta_\perp^{ik}\delta_\perp^{j\ell}+
\delta_\perp^{i\ell}\delta_\perp^{jk}}{(n-2)n}
\int_{\bm{q}}
\bm{q}_\perp^4
\,h(\bm{q},\bm{a}_1,\bm{a}_2).
\end{eqnarray}

The tensor-integral-reduction strategy described in this section 
is exhaustive for any tensor integral of 
arbitrary rank 
in the $n$-dimensional Euclidean space.
This can equally be applied to the $d$-dimensional Minkowski space
that is summarized in Appendix~\ref{app:cov}.

\section{Summary\label{sec:sum}}
Symmetries of a physical system 
provide us with a quite powerful
tool to greatly simplify theoretical
calculations of physical observables. 
The totally symmetric isotropic
tensor which is the generalized version of the Kronecker delta of
rank-2 tensor appears in various applications of particle physics, 
molecular dynamics, fluid dynamics, material sciences, and so forth. 
Thus, it is essential to know exact formulas 
of those tensors
to calculate angle averages
of an isotropic system and the corresponding value for a system that 
has partial isotropies in subspaces.

We have illustrated a rigorous approach to derive the
totally symmetric isotropic tensor 
$\widetilde{\mathcal{I}}^{i_1\cdots i_k}_{(k)}$
of arbitrary rank $k$ in
the $n$-dimensional Euclidean space. 
The derivation is based on only the abstract algebraic 
structure and symmetries.
All of the tensors of rank odd vanish and 
$\widetilde{\mathcal{I}}^{i_1\cdots i_{2k}}_{(2k)}$
is expressed as a linear combination of products of Kronecker
deltas as shown in Eq.~(\ref{eq:N-2k-summation}).
The approach has been generalized to analyze a physical system
that has totally symmetric isotropic components in a subspace.
As an immediate application, we have demonstrated that
angle averages can be evaluated without carrying out cumbersome
angular integration.
Instead, the symmetric properties of the tensor
$\widetilde{\mathcal{I}}^{i_1\cdots i_{2k}}_{(2k)}$
enable us to determine the averages by only counting combinatoric
multiplicity factors.

Loop integrals appearing in
 perturbative calculations of the gauge-field theory
may have divergences. To carry out a standard renormalization procedure, 
one first regularizes such an integral using
dimensional regularization. After imposing dimensional regularization,
there are numerous tensor integrals in $n=3-2\epsilon$ spatial dimensions.
We have demonstrated a systematic procedure to reduce
those tensor integrals as a linear combination of constant tensors
whose coefficients are scalar integrals. This straightforward demonstration 
of the tensor-integral reduction in the
 $n$-dimensional Euclidean space 
is equally applicable to the Minkovski space problems
such as the Passarino-Veltman reduction without losing generality.

\appendix
\section{Spherical polar coordinates in $\bm{n}$ dimensions
\label{app:polar-azimuthal}}
In this appendix, we illustrate the standard parametrization of
the spherical polar coordinates and
the corresponding solid-angle element in $n$ dimensions.
\subsection{Polar and azimuthal angles}
The spherical polar coordinates consist of 
the radius $r$, $n-2$ polar angles $\theta_i$,
and a single azimuthal angle $\phi$,
where $i=1$, $\cdots$, $n-2$. The construction of this system
can be easily achieved by applying the Pythagoras theorem 
recursively.

The radial vector $\bm{r}$ can be
expressed in terms of the Cartesian coordinates as
\begin{equation}
\bm{r}=(x^1,\cdots,x^n).
\end{equation}
Its magnitude is the radius, 
$r=\sqrt{\bm{r}^2}=\sqrt{x^kx^k}$, and the unit
radial vector $\hat{\bm{r}}\equiv\bm{r}/r$ is a function of
polar and azimuthal angles. 
If $n=2$, then 
the constraint
$\hat{\bm{r}}^2=(\hat{r}^{1})^2+(\hat{r}^{2})^2=1$
allows us to parametrize the coordinates as
$\hat{r}^{1}=\sin\phi$ and
$\hat{r}^{2}=\cos\phi$.
Similarly, one can parametrize
the last two coordinates for $n\ge 3$ as
$\hat{r}^{n}=\sqrt{1-\sum_{k=1}^{n-2}(\hat{r}^k)^2}\cos\phi$ and 
$\hat{r}^{n-1}=\sqrt{1-\sum_{k=1}^{n-2}(\hat{r}^k)^2}\sin\phi$.
Thus we require a single azimuthal angle $\phi$
for all $n\ge 2$ with the allowed range $0\le \phi\le 2\pi$.
The first $n-2$ coordinates 
$\hat{r}^1$, $\cdots$, $\hat{r}^{n-2}$
for $n\ge 3$ are parametrized by only polar angles as follows:
We set $\hat{\bm{\gamma}}_1\equiv\hat{\bm{r}}$.
We can always decompose $\hat{\bm{\gamma}}_1$ into
$\hat{\bm{\gamma}}_1=\bm{\alpha}_1+\bm{\beta}_1$, where
$\bm{\alpha}_1\equiv(\hat{\gamma}_1^1,0,\cdots,0)$ and 
$\bm{\beta}_1\equiv(0,\hat{\gamma}_1^2,\cdots,\hat{\gamma}_1^n)$ are 
along and perpendicular to the $x^1$ axis, respectively.
Because $\hat{\bm{\gamma}}_1^2=\bm{\alpha}_1^2+%
\bm{\beta}^2_1=1$, 
we can introduce a polar angle $\theta_1$
such that $
-1\le\hat{r}^1=\hat{\gamma}_1^1=\cos\theta_1\le 1$ and  
$0\le|\bm{\beta}_1|=
\sqrt{\sum_{i=2}^{n}(\hat{\gamma}_1^i)^2}
=\sin\theta_1\le 1$.
The allowed range of the polar angle 
$\theta_1$ is $0\le \theta_1\le \pi$. 
We can define the unit vector
$\hat{\bm{\gamma}}_k\equiv\bm{\beta}_{k-1}/\sin\theta_{k-1}$
recursively
for $k=2$, $\cdots$, $n-2$.
In a similar manner, we can decompose
$\hat{\bm{\gamma}}_k$ into
$\hat{\bm{\gamma}}_k=\bm{\alpha}_k+\bm{\beta}_k$ with
$\bm{\alpha}_k\equiv(\hat{\gamma}_k^k,0,\cdots,0)$ and 
$\bm{\beta}_k\equiv(0,\hat{\gamma}_k^{k+1},\cdots,\hat{\gamma}_k^n)$,
where we have neglected the $x^1$, $\cdots$, $x^{k-1}$ components
that are vanishing.
Because $\hat{\bm{\gamma}}_k^2=\bm{\alpha}_k^2+%
\bm{\beta}^2_k=1$, 
we can introduce a polar angle $\theta_k$
such that $-1\le\hat{\gamma}_k^k=\cos\theta_k\le 1$ and  
$0\le
|\bm{\beta}_k|=
\sqrt{\sum_{i={k+1}}^{n}(\hat{\gamma}_1^i)^2}=\sin\theta_k\le 1$, where
$0\le \theta_k\le \pi$.

In summary, the Cartesian coordinates for a unit radial
vector for $n\ge 2$
is parametrized with $n-2$ polar angles and an
azimuthal angle as
\begin{equation}
\label{complete-polar-coordinates}
\begin{array}{lll}
\hat{r}^{1}&=&\cos\theta_1,
\\
\hat{r}^{2}&=&\sin\theta_1\cos\theta_2,
\\
\hat{r}^{3}&=&\sin\theta_1\sin\theta_2\cos\theta_3,
\\
&\vdots&
\\
\hat{r}^{n-2}&=&\sin\theta_1\sin\theta_2\sin\theta_3\cdots
\sin\theta_{n-3}\cos\theta_{n-2},
\\
\hat{r}^{n-1}&=&\sin\theta_1\sin\theta_2\sin\theta_3\cdots
\sin\theta_{n-3}\sin\theta_{n-2}\sin\phi,
\\
\hat{r}^n&=&\sin\theta_1\sin\theta_2\sin\theta_3\cdots
\sin\theta_{n-3}\sin\theta_{n-2}\cos\phi.
\end{array}
\end{equation}
For $n=2$, only an azimuthal angle is required to express
$\hat{r}^1=\sin\phi$ and $\hat{r}^2=\cos\phi$.
\subsection{Solid angle}
\subsubsection{Gaussian-integral method}
The computation of the solid angle
$\Omega^{(n)}$ in the $n$-dimensional Euclidean space
can be carried out by making use of a Gaussian integral:
\begin{equation}
I^{(1)}\equiv\int_{-\infty}^\infty\!dx\, e^{-x^2}=\sqrt{\pi}.
\end{equation}
The $n$-dimensional Gaussian integral,
\begin{equation}
\label{eq:I1n}
I^{(n)}\equiv
\int_{-\infty}^\infty dx^1
\int_{-\infty}^\infty dx^2
\cdots
\int_{-\infty}^\infty dx^n\,
 e^{-[(x^1)^2+\cdots+(x^n)^2]},
\end{equation}
can be evaluated in the spherical polar
coordinate system in which the integrand is independent
of the direction of the Euclidean vector
$\bm{r}=(x^1,\cdots,x^n)$ whose radius is defined by
$r=\sqrt{(x^1)^2+\cdots+(x^n)^2}$.
The integrand of (\ref{eq:I1n}) depends only on $r$
and it is independent of the direction of $\bm{r}$.
Then the radial integral for $I^{(n)}$ 
is evaluated as
\begin{equation}
I^{(n)}=
\Omega^{(n)}
\int_{0}^\infty dr\,r^{n-1}
 e^{-r^2}=
\Omega^{(n)}\frac{\Gamma(\tfrac{1}{2}n)}{2},
\end{equation}
where the gamma function is defined by
\begin{equation}
\Gamma(x)\equiv\int_0^\infty t^{x-1}e^{-t}dt.
\end{equation}
Because $I^{(n)}$ is the $n$th power of $I^{(1)}$,
we have $I^{(n)}=\pi^{n/2}$.
This determines the solid angle $\Omega^{(n)}$ in $n$ dimensions
as
\begin{equation}
\label{Omega-n-formula}
\Omega^{(n)}=\frac{2\pi^{\frac{n}{2}}}{\Gamma(\tfrac{1}{2}n)}.
\end{equation}
\subsubsection{Angular-integral derivation}
The parametrization (\ref{complete-polar-coordinates})
can be used to express an $n$-dimensional volume integral
for $n\ge 2$
as a product of the radial integral and the angular one as:
\begin{eqnarray}
\label{eq:differential-solid-angle}
\int_{-\infty}^\infty\!\!\!dx^1
\int_{-\infty}^\infty\!\!\!dx^2\cdots
\int_{-\infty}^\infty\!\!\!dx^n
&=&\int_0^\infty dr\, r^{n-1}
\int_0^{2\pi}\!\!\!d\phi
\int_0^{\pi}\!\!\!d\theta_1
\int_0^{\pi}\!\!\!d\theta_2
\cdots
\int_0^{\pi}\!\!\!d\theta_{n-2}\,
J
\nonumber\\
&=&\int_0^\infty dr\, r^{n-1}
\int d\Omega_{\hat{\bm{r}}}^{(n)},
\end{eqnarray}
where $d\Omega_{\hat{\bm{r}}}^{(n)}$ and $J$ are 
the solid-angle element of
the unit radial vector $\hat{\bm{r}}$ 
and the Jacobian, respectively, 
and they are defined by 
\begin{equation}
\begin{array}{rll}
d\Omega_{\hat{\bm{r}}}^{(n)}
&=&
d\phi\, d\theta_1\,\cdots \,d\theta_{n-2}\,J,
\\
J&=&
\sin^{n-2}\theta_{1}
\sin^{n-3}\theta_{2}
\cdots
\sin^2\theta_{n-3}
\sin\theta_{n-2}.
\end{array}
\end{equation}
For $n=2$, there is no polar angle and $J=1$.
Changing the integration variables for polar angles from 
$\theta_i$ to 
$z_i\equiv \cos\theta_i$, we find that
\begin{equation}
\label{dOmegan}
\begin{array}{rll}
d\Omega_{\hat{\bm{r}}}^{(n)}
&=&
d\phi \, dz_1\,\cdots \,dz_{n-2}\,J_z,
\\
J_z&=&
(1-z_1^2)^{\frac{n-3}{2}}
(1-z_2^2)^{\frac{n-4}{2}}
\cdots
(1-z_{n-3}^2)^{\frac{1}{2}}\cdot 1.
\end{array}
\end{equation}
Note that $\sin\theta_i=(1-z_i^2)^{\frac{1}{2}}\ge0$
because $0\le \theta_i \le \pi$.
By integrating over $\phi$ and $z_i$'s,
we can reproduce 
the solid-angle formula (\ref{Omega-n-formula}):
\begin{eqnarray}
\label{eq:solid-angle-element}
\Omega^{(n)}
&=&\int d\Omega^{(n)}_{\hat{\bm{r}}}
\nonumber\\
&=&
\int_0^{2\pi} d\phi\,
\prod_{j=1}^{n-2}
\int_{-1}^1 dz_j(1-z_j^2)^{\tfrac{n-2-j}{2}}
\nonumber\\
&=& 2\pi 
\frac{\Gamma(\tfrac{1}{2}n-\tfrac{1}{2})\Gamma(\tfrac{1}{2})}
{\Gamma(\tfrac{1}{2}n)}
\frac{\Gamma(\tfrac{1}{2}n-\tfrac{3}{2})\Gamma(\tfrac{1}{2})}
{\Gamma(\tfrac{1}{2}n-\tfrac{1}{2})}
\cdots
\frac{\Gamma(\tfrac{3}{2})\Gamma(\tfrac{1}{2})}
{\Gamma(\tfrac{5}{2})}
\frac{\Gamma(1)\Gamma(\tfrac{1}{2})}
{\Gamma(\tfrac{3}{2})}
\nonumber\\
&=&\frac{2\pi^{\frac{n}{2}}}{\Gamma(\tfrac{1}{2}n)},
\end{eqnarray}
where we have used $\Gamma(\tfrac{1}{2})=\sqrt{\pi}$ 
and the integral table
\begin{equation}
\int_{-1}^1 (1-x^2)^{n}dx
=\frac{\Gamma(n+1)\Gamma(\tfrac{1}{2})}{\Gamma(n+\tfrac{3}{2})}.
\end{equation}

\section{Direct evaluation of angle average
\label{app:explicit-evaluation-adotkton}}
By making use of the definition for the tensor angular integral in 
Eq.~(\ref{isotropic-angular-integral}), we can express
the average of $(\bm{a}\cdot\hat{\bm{r}})^k$ over
the direction of the unit radial vector
$\hat{\bm{r}}$ as
\begin{equation}
\langle (\bm{a}\cdot\hat{\bm{r}})^k\rangle_{\hat{\bm{r}}}
=\frac{1}{\Omega^{(n)}}
\int d\Omega^{(n)}_{\hat{\bm{r}}}(\bm{a}\cdot\hat{\bm{r}})^k,
\end{equation}
where $k$ is a non-negative integer
and $\bm{a}$ is a constant vector.
The solid-angle element $d\Omega_{\hat{\bm{r}}}^{(n)}$
and the $n$-dimensional solid angle $\Omega^{(n)}$ are defined in
Eqs.~(\ref{dOmegan}) and
(\ref{eq:solid-angle-element}), respectively.
Because the integrand $(\bm{a}\cdot\hat{\bm{r}})^k$ is a scalar,
the average is invariant under rotation. Thus, there exists 
a rotational transformation to make $x^1$ Cartesian axis 
parallel to $\bm{a}$ so that $\bm{a}=(|\bm{a}|,0,\cdots,0)$.
 In that case, the average is simplified as
\begin{equation}
\langle (\bm{a}\cdot\hat{\bm{r}})^k\rangle
=
|\bm{a}|^k
\langle (\hat{r}^1)^k\rangle.
\end{equation}
By making use of the parametrization for $d\Omega_{\hat{\bm{r}}}^{(n)}$
in Eq.~(\ref{dOmegan}) and integrating over $\phi$, $z_2$, $\cdots$,
$z_{n-2}$, we find that
\begin{equation}
\langle (\hat{r}^1)^k\rangle
=
\frac{\int_{-1}^1 dz_1 z_1^{k} (1-z_1^2)^{\frac{1}{2}(n-3)}}
 {\int_{-1}^1 dz_1 (1-z_1^2)^{\frac{1}{2}(n-3)}}.
\end{equation}
For all $k$ odd, the integrand of the numerator is odd to make the
integral vanish.
By making use of the integral table,
\begin{equation}
\label{integral-table-angle-average}
\int_{-1}^1 (1-x^2)^a x^{b}dx
=
\frac{1+(-1)^{b}}{2}
\frac{\Gamma(1+a)\Gamma(\frac{1}{2}+\frac{1}{2}b)}
{\Gamma(a+\frac{1}{2}b+\frac{3}{2})},
\end{equation}
and the identities
\begin{equation}
\begin{array}{rll}
\displaystyle
\frac{\Gamma(\frac{k+1}{2})}{\Gamma(\frac{1}{2})}&=&
\displaystyle\frac{(k-1)!!}{\sqrt{2^k}},
\\[3ex]
\displaystyle
\frac{\sqrt{2^k} \Gamma(\tfrac{n+k}{2})}{\Gamma(\tfrac{n}{2})}
&=&
n(n+2)\cdots(n+k-4)(n+k-2),
\end{array}
\end{equation}
we find that
\begin{equation}
\label{eq:ar-2k-angle-result}
\langle (\bm{a}\cdot\hat{\bm{r}})^{k}\rangle=
\begin{cases}
0,&k\,\,\textrm{odd},
\\
\displaystyle |\bm{a}|^k
\frac{\Gamma(\frac{n}{2})\Gamma(\frac{k+1}{2})}{
\sqrt{\pi}\Gamma(\frac{n+k}{2})}
=
\frac{|\bm{a}|^k(k-1)!!}{n(n+2)(n+4)\cdots(n+k-2)},& k\,\,\textrm{even}.
\end{cases}
\end{equation}
This result for the angular integral agrees with  
Eq.~(\ref{eq:ar-2k-wo-angle-result}) that is obtained by 
taking into account the symmetries only.

The most general form of the angle average is 
\begin{equation}
\langle
(\bm{a}_1\cdot\hat{\bm{r}})
(\bm{a}_2\cdot\hat{\bm{r}})
\cdots
(\bm{a}_m\cdot\hat{\bm{r}})
\rangle_{\hat{\bm{r}}}
=
\frac{1}{\Omega^{(n)}}
\int
d\Omega^{(n)}_{\hat{\bm{r}}}
(\bm{a}_1\cdot\hat{\bm{r}})
(\bm{a}_2\cdot\hat{\bm{r}})
\cdots
(\bm{a}_m\cdot\hat{\bm{r}}),
\end{equation}
where all of the constant vectors $\bm{a}_i$ do not have to be
distinct and $m$ is a positive integer which is not bounded
above.
The integral can be expressed as a linear combination
of
\begin{equation}
\langle
(\hat{r}^1)^{k_1}
\cdots
(\hat{r}^n)^{k_n}
\rangle_{\hat{\bm{r}}}
=
\frac{\Gamma(\tfrac{1}{2}n)}{2\pi^{n/2}}
\int_0^{2\pi}\!\!d\phi\,\sin^{k_{n-1}}\phi\,\cos^{k_{n}}\phi
\prod_{j=1}^{n-2}
\int_{-1}^1 dz_j 
z_j^{k_j} 
(1-z_j^2)^{\frac{1}{2}(n-2-j+\sum_{\ell=j+1}^n k_\ell)},
\end{equation}
where $k_i$'s are non-negative integers satisfying
$\sum_{i=1}^n k_i=m$. 
In principle, the evaluation of the integral is straightforward.
However, the corresponding calculations consist of many steps and
one should take extreme care to avoid mistakes during such a
tedious computation.
Instead, by making use of the totally symmetric
isotropic tensor given in
Eq.~(\ref{eq:ar-general-result}), one can greatly reduce the efforts
and obtain the result only by counting combinatoric multiplicity
factors.

\section{Reduction in the $\bm{d}$-dimensional 
Minkowski space\label{app:cov}}
In evaluating divergent loop integrals coming from perturbative
calculations of the gauge-field theory,
one regularizes those integrals by analytic continuation of
the space-time dimensions from 4 to $d=4-2\epsilon$ to express
the integral measure as
\begin{equation}
\int_q\equiv\int d^dq=
\int_{-\infty}^\infty dq^0
\int_{-\infty}^\infty dq^1
\cdots
\int_{-\infty}^\infty dq^{d-1},
\end{equation}
where the contravariant $d$-vector $q^\mu=(q^0,q^1,\cdots,q^{d-1})$ 
is the loop momentum.
Here, $q^0$ and $q^i$ correspond to the time and spatial 
components, respectively. Note that a Greek letter is used
for a $d$-vector index in a Minkowski space while an italic index
is used for the Euclidean space.

One can integrate out $q^0$ by closing the contour
on the complex $q^0$ plane to find the residue 
originated from the relevant propagator factors.
Then the integral reduces into a form defined in  
the $(d-1)$-dimensional Euclidean space that can be always evaluated
by making use of formulas presented in the text.
Alternatively, without carrying out the $q^0$ integral first, 
one can directly evaluate the $d$-dimensional integral as follows.

The scalar product of two $d$-vectors $a$ and $b$, 
which is invariant under Lorentz transformation, is defined by
\begin{equation}
a\cdot b=a^\mu b^\nu g_{\mu\nu}=a^0 b^0-a^1b^1-\cdots-a^{d-1}b^{d-1},
\end{equation}
where $g_{\mu\nu}=g^{\mu\nu}=\textrm{diag}[1,-1,-1,\cdots,-1]$
is the metric tensor for the $d$-dimensional Minkowski space
that corresponds to 
the Kronecker delta $\delta^{ij}$ in the Euclidean space.
The covariant $d$-vector $a_\mu$ corresponding to
the contravariant $d$-vector $a^\mu=(a^0,a^1,a^2,\cdots, a^{d-1})$
is defined by $a_\mu=g_{\mu\nu}a^\nu=(a^0,-a^1,-a^2,\cdots, -a^{d-1})$
and $a^\mu=g^{\mu\nu}a_\nu$.

By generalizing Eq.~(\ref{eq:tensor-A-reduced}) to 
the $d$-dimensional Minkowski space, 
we can reduce 
the tensor loop integral that depends only on the loop momentum $q$
into the following form
\begin{equation}
\int_q q^{\mu_1}\cdots q^{\mu_{2k}} f(q)
=\widetilde{\mathcal{I}}^{\mu_1\mu_2\cdots \mu_{2k}}_{(2k)}
\int_q (q\cdot q)^k f(q),
\end{equation}
where $f(q)$ is a Lorentz scalar
which is invariant under Lorentz transformation and
we have neglected the vanishing contributions
of rank odd.
The totally symmetric isotropic tensor
of rank $2k$,
\begin{equation}
\label{eq:N-2k-summation-mu}
\widetilde{\mathcal{I}}_{(2k)}^{\mu_1\mu_2\cdots \mu_{2k-1}\mu_{2k}}
=
\frac{1}{d(d+2)\cdots (d+2k-2)}
\frac{1}{(2k)!!}
\sum_{\sigma}
g^{\sigma(\mu_1)\sigma(\mu_2)}\cdots
g^{\sigma(\mu_{2k-1})\sigma(\mu_{2k})},
\end{equation}
is the generalized version into the $d$-dimensional 
Minkowski space: $n$ and $\delta^{ij}$ in Eq.~(\ref{eq:N-2k-summation})
are replaced with $d$ and $g^{\mu\nu}$, respectively.

In a similar manner, we can reduce
the tensor loop integral that depends on the loop momentum $q$ and an
external momentum $a$  for a massive particle $(a\cdot a>0)$ as
\begin{eqnarray}
\label{eq:tensor-minkow-external-1}
\int_q q^{\mu_1}\cdots q^{\mu_{k}} f(q,a)
&=&
\sum_{r=0}^{k}
\frac{1}{r!(k-r)!}
\sum_{\sigma}a^{\sigma(\mu_1)}\cdots a^{\sigma(\mu_{r})}
\,\widetilde{\mathcal{I}}_{\perp(k-r)}
^{\sigma(\mu_{r+1})\cdots \sigma(\mu_{k})}
\nonumber \\
&&\times
\int_q \frac{(a\cdot q)^{r}(q_\perp\cdot q_\perp)^{(k-r)/2}}{(a\cdot a)^{r}} f(q,a),
\end{eqnarray}
where the summation over $r$ is for even $k-r$ only.
The totally symmetric isotropic tensor
of rank even is given by
\begin{equation}
\label{eq:N-2k-summation-mu}
\widetilde{\mathcal{I}}_{\perp(2k)}^{\mu_1\mu_2\cdots \mu_{2k-1}\mu_{2k}}
=
\frac{1}{(d-1)(d+1)\cdots (d+2k-3)}
\frac{1}{(2k)!!}
\sum_{\sigma}
g_\perp^{\sigma(\mu_1)\sigma(\mu_2)}\cdots
g_\perp^{\sigma(\mu_{2k-1})\sigma(\mu_{2k})}.
\end{equation}
Here, $q_\perp^\mu=g_\perp^{\mu\nu} q_\nu$ and
$g_\perp^{\mu\nu}$ is defined by
\begin{equation}
g_\perp^{\mu\nu}
=
g^{\mu\nu}-\frac{a^\mu a^\nu}{a\cdot a}.
\end{equation}
If the scalar function $f(q,a)$ is replaced with 
$f(q,a_1,\cdots, a_m)$, where $a_i$ is the $d$-momentum
of the $i$th external massive particle, then one can generalize
the method to obtain Eq.~(\ref{tensor-C}) in a similar manner
that we have employed to derive the relativistic version in 
Eq.~(\ref{eq:tensor-minkow-external-1})
in the presence of a single external particle.


\begin{acknowledgments}
We thank Soo-hyeon Nam and Chaehyun Yu
for their careful reading the manuscript and useful comments.
The work of J.-H.E.\ was
supported by Global Ph.D. Fellowship Program through the National
Research Foundation (NRF) of Korea funded by the Korea government (MOE)
under Contract No.\ NRF-2012H1A2A1003138.
The work of D.-W.J.\ was supported by NRF under
Contract No. NRF-2015R1D1A1A01059141.
This work was supported by the Do-Yak project of
NRF under Contract No.\ NRF-2015R1A2A1A15054533.
\end{acknowledgments}

\end{document}